\documentclass[superscriptaddress,twocolumn,showpacs,amsmath,amssymb,aps,prl,floatfix]{revtex4-1}
\usepackage[latin1]{inputenc}
\usepackage[american]{babel}
\usepackage[T1]{fontenc}
\usepackage{mathrsfs}
\usepackage{hyperref}
\usepackage{multirow}
\usepackage{bm, epsfig}
\usepackage{graphicx}
\usepackage{subeqnarray}
\usepackage{bbold}
\usepackage{upgreek}

\hypersetup{colorlinks=true, urlcolor=blue, citecolor=blue, filecolor=blue, linkcolor=blue}

\newcommand{\nemc}{NE$\mu$C}

\begin{document}

{\bf Comment on ``Nuclear Excitation by Free Muon Capture"}

\vspace{5mm}

In the paper \cite{Gargiulo_PRL2023} the process of free muon capture with simultaneous excitation of a nuclear isomer has been suggested, claiming that ``the effect can be detectable for selected isotopes''.  
Here, we argue that this claim can not be confirmed.
Briefly, the process is far from the dominant mechanism for nuclear excitation; it excites high energy nuclear levels that will not generally decay to the isomer; the proposal assumes all incident muons will fulfil energy criteria, ignoring dominant capture paths; and nuclei excited by muons will have a shortened lifetime due to muonic capture.

\begin{figure}[b!]
    \centering
    \includegraphics[width=0.9\columnwidth]{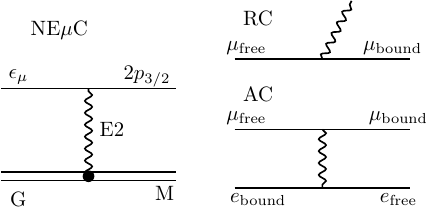}
    \caption{Competing capture modes of the muon in the continuum. NE$\mu$C is only available with matching energy and angular momentum. By contrast, radiative (RC) and Auger capture (AC) rates are much larger and final states are always available.}
    \label{fig:rates}
\end{figure}

Let us start by discussing an important technical point. As stressed in \cite{Gargiulo_PRL2023}, for a free muon to be captured to its ground or first excited state, it should have a well-defined energy close to the nuclear resonance. Coupled with the small size of the target orbital, such low energy, non-relativistic scattering will be dominated by the $s$-wave cross-section, however this is in conflict with angular momentum and parity selection rules for most of the considered transitions in Table I. 
Instead these must originate from $p$ or $d$-wave muons, where the rate will be suppressed. On the other hand, radiative and Auger capture via dominant channels are always available (see Fig.~\ref{fig:rates})  and can involve any free muon energies and bound muon states.

After a muon is captured in a highly excited state with a statistically distributed angular momentum, the dominant process is cascade towards the ground state, first via Auger, then via radiative decay~\cite{Okumura_PRL2021}. Due to the similar energy scale for muonic and nuclear states, there is strong mixing of muon-nucleus levels driven by hyperfine interaction. This so-called dynamical splitting was discussed in \cite{Wu_1969, BorieRinker1982} and improved later in \cite{Michel_PRA2019}. The correctness of the theoretical prediction was fully confirmed by the experiment \cite{Mux_PRC2020}. As a result, all states of the muonic cascade include a superposition of a few low-lying nuclear states. Decays in the cascade occur spontaneously via photon emission, without the requirement of precise energy matching, and nuclear excitations are merely a side effect of this cascade in muonic atoms. This is a vastly different physical reality from that presented in \citep{Gargiulo_PRL2023}, where the muonic and nuclear degrees of freedom are considered highly separable.

A major drawback of the paper is that it compares the probability of nuclear excitation by muon capture (\nemc) almost exclusively with that of nuclear excitation by electron capture (NEEC). These are two very different physical systems: NEEC occurs with no competing nuclear excitation mechanisms. On the other hand, in order to properly evaluate the experimental feasibility of the proposal, the \nemc\ process should be compared with other decay channels of the same system to establish the hierarchy. The dominant process for muonic atoms, namely excitation upon muon cascade, has not received enough attention in the paper \citep{Gargiulo_PRL2023}.

Even if one allows that the \nemc\ mechanism might occur in some systems as a sub-dominant effect, the nuclear levels that are excited by this method are relatively high-energy, and do not necessarily cascade to the metastable isomer. For instance, the suggested nuclear state of $^{207}$Pb at 4980.5 keV is highly excited and the direct photo-excitation rate is low. However,  the excited state is separated from the ground state by over 100 levels and it is not metastable, so it would uncontrollably decay in gamma cascade. Therefore, the \nemc\ process will not enable the preferential feeding of nuclear isomers.

Finally, excited nuclei produced by NE$\mu$C will generally be destroyed by nuclear muon capture, which is the dominant decay mechanism for muonic atoms with heavy nuclei \cite{BorieRinker1982}.
Overall, the paper \cite{Gargiulo_PRL2023} presents an interesting mechanism for manipulating nuclear states via interaction with a muon, but ignores the dominant mechanism of muon-nucleus interaction, namely the muonic dynamical-structure cascade, and  gives a misleading impression that \nemc\ could be observed. Despite the idea's attractiveness, based on the points mentioned above, it is highly improbable that \nemc\ can be visible in an experiment.

N.S.O. thanks the Gordon Godfrey fund for the financial support of the visit to UNSW Sydney, Australia. 

\vspace{5mm}

Natalia S. Oreshkina$^1$ and Julian C. Berengut$^2$ \\
{\small $^1$ Max-Planck-Institut f\"{u}r Kernphysik, Saupfercheckweg 1, 69117 Heidelberg, Germany \\
$^2$ School of Physics, University of New South Wales, Sydney NSW 2052, Australia}


\bibliography{refs}

\end{document}